%% file: APL-accepted-arxiv.tex
\documentclass[aip,amsmath,amssymb,reprint]{revtex4-1}

\usepackage[T1]{fontenc}
\usepackage{lmodern}
\usepackage{txfonts}

\usepackage{fancyhdr}
\fancypagestyle{append}{%
        \fancyhead[LE,RO]{\thepage}
        \fancyhead[LO,Re]{\textit{Light and heavy hole diffusion},
         ~~~Alexandre Walker and Mike Denhoff}
          \fancyfoot[C]{Appendix, Oct 2017}
           }
\fancypagestyle{plain}{%
        \fancyhead[LE,RO]{}
        \fancyhead[LO,Re]{}
          \fancyfoot[C]{Preprint, Oct 2017}
           }

\usepackage{graphicx}
\usepackage{dcolumn}
\usepackage{bm}
\usepackage{textcomp}
 
\graphicspath{{pic/}}

\usepackage{listings}
\usepackage{verbatim}

\usepackage[colorlinks=true, allcolors=blue]{hyperref}

\begin{document} 

\preprint{AIP/123}


\title{Heavy and light hole minority carrier transport properties in low-doped n-InGaAs lattice matched to InP}

\author{Alexandre W. Walker}
\author{Mike W. Denhoff}
\affiliation{National Research Council of Canada, 1200 Montreal Road,
            M-50, K1A 0R6, Ottawa, Ontario, Canada\\[1ex]
            This is the ``accepted manuscript'' of the article published as \\
    Alexandre W. Walker and Mike W. Denhoff, 
   Appl. Phys. Lett. 111, 162107 (2017); doi: 10.1063/1.5002677\\
   The appendix is not part of the Appl. Phys. Lett. article.}

\date{31 October 2017}
\email{alexandre.walker@nrc-cnrc.gc.ca}

 
\keywords{InGaAs, minority carrier diffusion length, mobility, lifetime, heavy hole, light hole}

\begin{abstract}
Minority carrier diffusion lengths in low-doped n-InGaAs using InP/InGaAs double-heterostructures are reported using a simple electrical technique.
 The contributions from heavy and light holes are also extracted using this methodology, including minority carrier mobilities and lifetimes.
 Heavy holes are shown to initially dominate the transport due to their higher valence band density of states, but at large diffusion distances, the light holes begin to dominate due to their larger diffusion length.
 It is found that heavy holes have a diffusion length of $54.5\pm0.6\, $\textmu m for an n-InGaAs doping of $8.4\times10^{15}\, \text{cm}^{-3}$ at room temperature, whereas light holes have a diffusion length in excess of $140\, $\textmu m.
 Heavy holes demonstrate a mobility of $692\pm63\,\text{cm}^{2}/\text{Vs}$ and a lifetime of $1.7\pm0.2\,$\textmu s, whereas light holes demonstrate a mobility of $6200\pm960\, \text{cm}^{-2}/\text{Vs}$ and a slightly longer lifetime of $2.6\pm1.0\,$\textmu s. The presented method, which is limited to low injection conditions, is capable of accurately resolving minority carrier transport properties.
\end{abstract}

\maketitle
\thispagestyle{plain}

\twocolumngrid

Minority carrier diffusion lengths are critically influential in optoelectronic device performance, including solar cells, photodetectors and heterojunction bipolar transistors for example. Several methods presently exist to measure these, including electron beam induced current (EBIC) along a cross-sectional \textit{pn} junction,\cite{Maximenko} cathodoluminescence (CL) measurements on double-heterostructures,\cite{Schultes, Gustafsson, Niemeyer} zero-field time-of-flight\cite{Sharma,Lovejoy}, surface phovotoltage\cite{Scroder, Kronik} to name a few. Coupling these diffusion lengths to lifetime measurements typically obtained from time-resolved photoluminescence, time-of-flight experiments, or time-resolved CL measurements,\cite{Boulou} one can then infer minority carrier mobilities if both measurements are performed for a comparable excess carrier concentration. These parameters can be critical in designing minority carrier devices, for example in cases where minority carrier mobilities exceed that of majority carrier mobilities\cite{Lovejoy} or in solar cells where minority carrier diffusion lengths are on the order of the active region thickness.\cite{Walkera}
   \begin{figure} [b]
   \begin{center}
   \begin{tabular}{cc} 
   \includegraphics[width=0.48\textwidth]{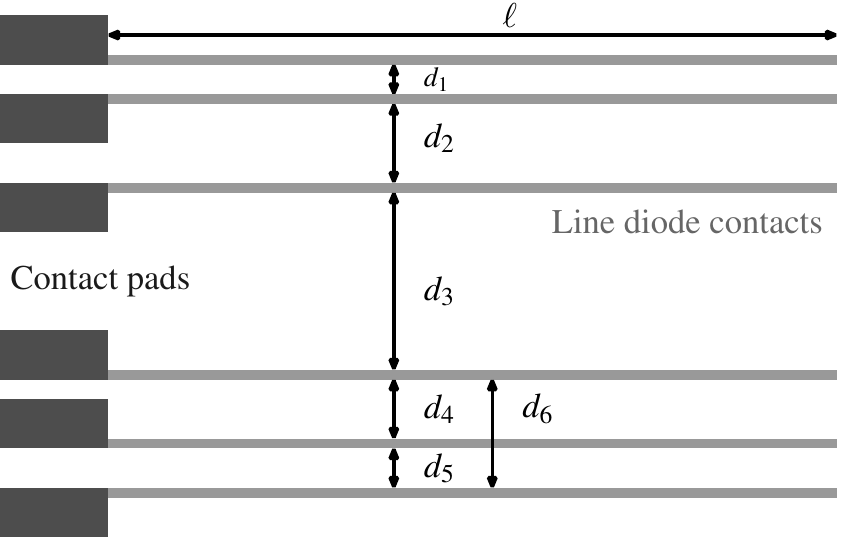}
   \end{tabular}
   \end{center}
   \caption[example]
   { \label{fig:fig-1}
 Set of long and thin diffused junction diodes with inter-diode spacings $d_{1..6}$, where the
diode length $\ell >> d_{1..6}$.}
   \end{figure}

Each of the aforementioned methodologies have their inherent advantanges and disadvantages.
 For example, the zero-field time-of-flight method is simple in principle, but becomes obfuscated when fitting the transient photovoltage for a heterojunction such as an InP/InGaAs photodetector.
 For cross-sectional EBIC, the method can be destructive since samples must be well polished for a clean cross-section, and the requirement of an electron microscope can result in significant time and effort.
 Thus it is of interest to develop an overall simpler method to measure minority carrier diffusion lengths. Here, we present a purely electrical method capable of extracting diffusion lengths that is effectively a steady-state Haynes Shockley experiment.\cite{Haynes, Shockley}
 The method is demonstrated using an n-type InP:Si/intrinsic InGaAs/n-type InP:Si double-heterostructure which is used to make \textit{pin} short wave infrared detectors and focal plane arrays.\cite{Walkerb}
 Furthermore, this method is shown to be capable of separating the contributions of heavy and light holes, and, with a few assumptions, determining the lifetimes and diffusion constants (i.e. mobilities).
 The required test devices can also be included on production wafers to monitor the material and fabrication quality across the wafer and from epitaxial run to run.
 These test structures consist of long ($750\,$\textmu m) and thin ($10\,$\textmu m) diffused junction line diodes as shown in Fig.~\ref{fig:fig-1}.
 The junction is created by Zn diffusion through the InP and about $100$ nm into the InGaAs material.
 The inter-line diode separations range from much shorter to much longer than the diffusion length.
 Note that, during the measurement, one can bypass one diode to obtain a larger distance, as illustrated by $d_6$ in Fig.~\ref{fig:fig-1}.

   \begin{figure} [t]
   \begin{center}
   \includegraphics[width=0.48\textwidth]{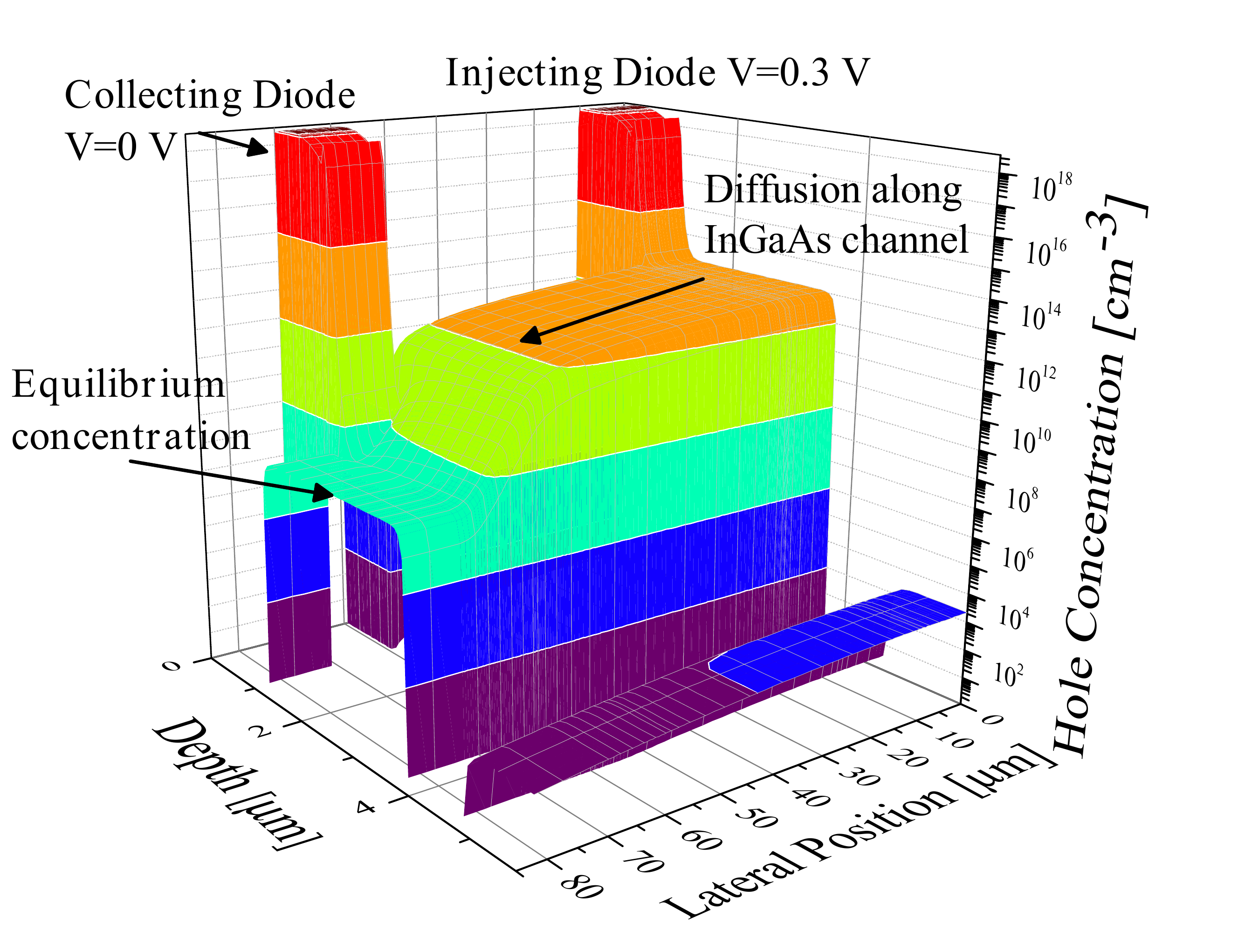}
   \end{center}
   \caption[example]
   { \label{fig:fig-2}
Simulated carrier concentration as a function of depth in device and lateral distance between two adjacent diodes using Atlas (v5.23.12.C) by Silvaco. Regions of very low hole concentration correspond to InP. The red to violet color coding represents the magnitude of the carrier concentration.}
   \end{figure}

The basic physics is summarized in detail elsewhere.\cite{Walkerb, Sze} Essentially, forward biasing one line diode injects minority carriers into the InGaAs layer leading to diffusion of carriers through the InGaAs channel. This can be seen in Fig.~\ref{fig:fig-2} showing a simulated cross-section of the device in terms of hole concentration as a function of depth and lateral position between two adjacent line diodes using Atlas by Silvaco (v.5.23.12.C), assuming default material properties.\cite{AtlasUG} The hole concentration is uniform across the thickness of the InGaAs layer since its thickness ($2.7\,$\textmu m) is significantly smaller than the diffusion length.
 The carriers which diffuse to an adjacent line diode are collected by applying a zero or slightly reverse bias with respect to the bottom InP layer which serves as the ground contact. Assuming low injection, if the length of the line diode $\ell$ is sufficiently long compared to the maximum interdiode distance $d_i$, and if $d_i$ is much longer than the InGaAs thickness, the hole current can be described by the one dimensional diffusion equation. The hole density under the injecting diode is $p_{\text{inj}}$, which is dictated by the applied bias, and the hole concentration at the collecting diode is the equilibrium value $p_0$ since the collecting diode is maintained at zero applied bias. Using these boundary conditions, and assuming no interface recombination, solving the diffusion equation leads to the solution in terms of the collected hole current density $J_p(W)$ as a function of interdiode separation $W$ given as\cite{Walkerb,Sze}
\begin{equation}
\label{eq:one}
J_p(W) = \frac{qD(p_{\text{inj}}-p_0)}{L\sinh(W/L)}, \\
\end{equation}
where $L$ is the diffusion length, $D=\mu k_B T/q$ is the diffusion coefficient, $\mu$ is the mobility, $k_B$ is Boltzmann's constant, $T$ is temperature and $q$ is the electronic charge.
 In the case of low injection, the initial (injected) hole density $p_{\text{inj}}$ can be calculated using $p_{\text{inj}}=n_i^2/N_D\text{exp}\left(qV/k_BT\right)$), where $n_i$ is the intrinsic carrier concentration and $N_D=8.4\times10^{15}\, \text{cm}^{-3}$ is the donor concentration in the InGaAs extracted using CV measurements on $200\,$\textmu m diameter devices.
 Finally, $p_0$ is the equilibrium hole concentration given by $p_0=n_i^2/N_D$.
 Note the cross-sectional area used to scale the measured current to a current density is given by $A=t\times \ell$, where $t$ is the \mbox{InGaAs} thickness and $\ell$ is the length of the diode.
 Fitting Eq.~(\ref{eq:one}) to experimental data thus reveals the minority carrier diffusion length $L$ and the mobility $\mu$, thus giving access to the lifetime $\tau$.
 An electron effective mass of $0.043\,m_e$ is assumed for InGaAs, along with heavy and light hole effective masses of $0.46\,m_e$ and $0.047\,m_e$ respectively, and a room temperature bandgap of 0.734 eV.\cite{Vurgaftman}
 This gives an intrinsic carrier concentration of $n_i=6.14\times10^{11}\, \text{cm}^{-3}$ for room temperature (T=$296$K).

However, an important consideration that is easy to overlook is the interdiode separation $W$.
 The simulation results of Fig.~\ref{fig:fig-2} shows that the hole concentration is nearly constant under the injecting diode, but starts decreasing by $1-2\%$ within ${\sim}1\,$\textmu m of the inside edge of the diode.
 On the other hand, the hole concentration drops rapidly to the equilibrium concentration from the leading edge of the collecting diode.
 The separation of the diodes therefore starts a small distance inside the edge of the injecting diode and ends some distance past the leading edge of the collecting diode.
 This separation correction, $\delta$, to the interdiode separation $W$, defined from the inner edges of two diodes, is discussed later. 

\begin{figure} [t]
   \begin{center}
   \includegraphics[width=0.48\textwidth]{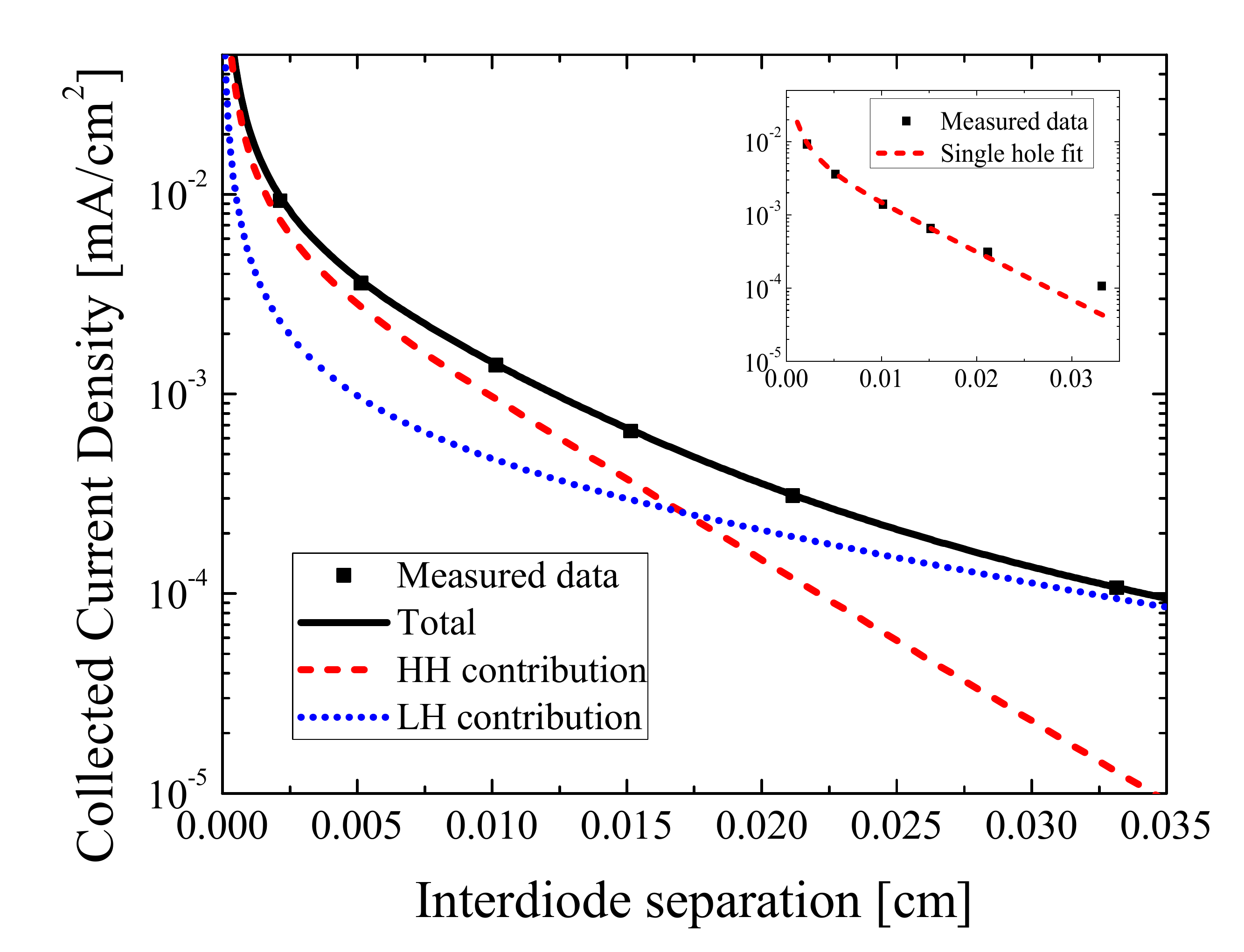}
   \end{center}
   \caption[example]
   { \label{fig:fig-3}
Measured collected current density as a function of interdiode separation
 $W$ for $V=0.3$V, including the best-fit to Eq. (\ref{eq:one}) for light
 and heavy holes and the sum of both terms.
Inset shows the fit according only to a single hole type.}
   \end{figure}

Current - voltage measurements were made using a Hewlett-Packard HP4155C Semiconductor Analyzer.
 Results from devices from two different wafers than the device reported here were reported in Ref.~\citenum{Walkerb} and are similar to the present results.
 A fit to measured collected current densities for an applied bias of 0.3V to Eq.~(\ref{eq:one}) with fitting parameters $L=66.6\,$\textmu m and $D=23.5\,\text{cm}^2/\text{s}$ is shown in the inset of Fig.~\ref{fig:fig-3}, assuming a separation correction of $\delta\,=\,2.5\,$\textmu m.
 The measured collected current for the largest separation of $330\,$\textmu m is well above the theoretical fit, and the point at $210\,$\textmu m is also above, albeit slightly.
 These higher than expected currents at large diffusion distances can be explained by the existance of light holes with a longer diffusion length than heavy holes. 

In order to account for both heavy and light holes, two versions of
 Eq.~(\ref{eq:one}), one for heavy holes and one for light holes, can be
 added together to give the total collected current.
 The number of fitting parameters doubles, which can lead to more ambiguity
 in the fit. One must first calculate the fraction of the injected hole
 concentration composed of heavy holes, $f_{HH}$, which is dictated by the
 ratio of heavy hole density of states to the total valence band density of
 states $N_v$ (ignoring the split-off band), given as
\begin{equation}
\label{eq:two}
f_{HH}=\frac{N_v^{HH}}{N_v^{HH}+N_v^{LH}} = \frac{m_{HH}^{3/2}}{m_{HH}^{3/2}+m_{LH}^{3/2}};
\end{equation}
the fraction for light hole occupation is analogous, or simply given
 as $f_{LH}=1-f_{HH}$. For the adopted heavy and light holes effective masses,
 heavy holes dominate the valence band density of states by a factor of ${\sim}30$. 

The best fit of Eq.~(\ref{eq:one}) for the sum of heavy and light holes, as well as their individual contributions, are illustrated in Fig.~\ref{fig:fig-3} for an applied bias of 0.3 V.
 The fit is achieved using a nonlinear least squares fitting algorithm from Matlab R2017a's curve fitting toolbox.
 The algorithm uses the diffusion length and diffusion constant for both light and heavy holes respectively as fitting parameters.
 Standard errors for the best-fit parameters are computed using the mean square errors and the Jacobian matrix at the solution (i.e. neglecting the higher order terms to approximate the Hessian matrix, which is reasonable since the residuals are close to zero at the solution).
 It is clear that adding a term specifically for the light holes yields a much better agreement for the largest interdiode separations.
 The heavy and light hole diffusion lengths are extracted as $53.9\pm0.8\,\mu \text{m}$ and $192\pm22\,\mu\text{m}$ respectively, again assuming a separation correction of $\delta\,=\,2.5\,$\textmu m.
 These can be compared to the diffusion length of $66.6\,\mu \text{m}$ found using the single carrier model.
 This single carrier value thus represents an effective diffusion length, combining both a shorter heavy hole diffusion length and a longer light hole diffusion length.
 These diffusion lengths are within the range of previously determined diffusion lengths at room temperature of $140\,$\textmu m for a lower doping of $10^{15}\, \text{cm}^{-3}$.\cite{Gallant}
 From the results in Fig.~\ref{fig:fig-3}, for short interdiode separations, the current is mainly due to heavy holes; in particular, for $W=0$, ${\sim}80\%$ of the current is due to heavy hole transport.
 This decreases to $50\%$ at $200\,$\textmu m, beyond which light holes dominate the transport.
 For minority carrier devices such as InGaAs/InP \textit{pnp} transistors, light holes can contribute ${\sim}20\%$ of the diffusion current.

It is important now to discuss the influence of the separation correction to the interdiode separation $W$ on the extracted diffusion length. For example, setting this to $\delta=0.5\,$\textmu m results in a heavy hole $L=58.6\pm3.5\,$\textmu m (a $9\%$ increase) and light hole $L=248\pm190\,$\textmu m (a $30\%$ increase); note the large uncertainties. Exceeding $2.5\,$\textmu m further decreases the diffusion lengths, but also increases the uncertainties. A value of $2.5\,$\textmu m is thus used in the remaining analysis, and is assumed to stay constant as a function of separation. 

  \begin{figure} [t]
   \begin{center}
    \includegraphics[width=0.48\textwidth]{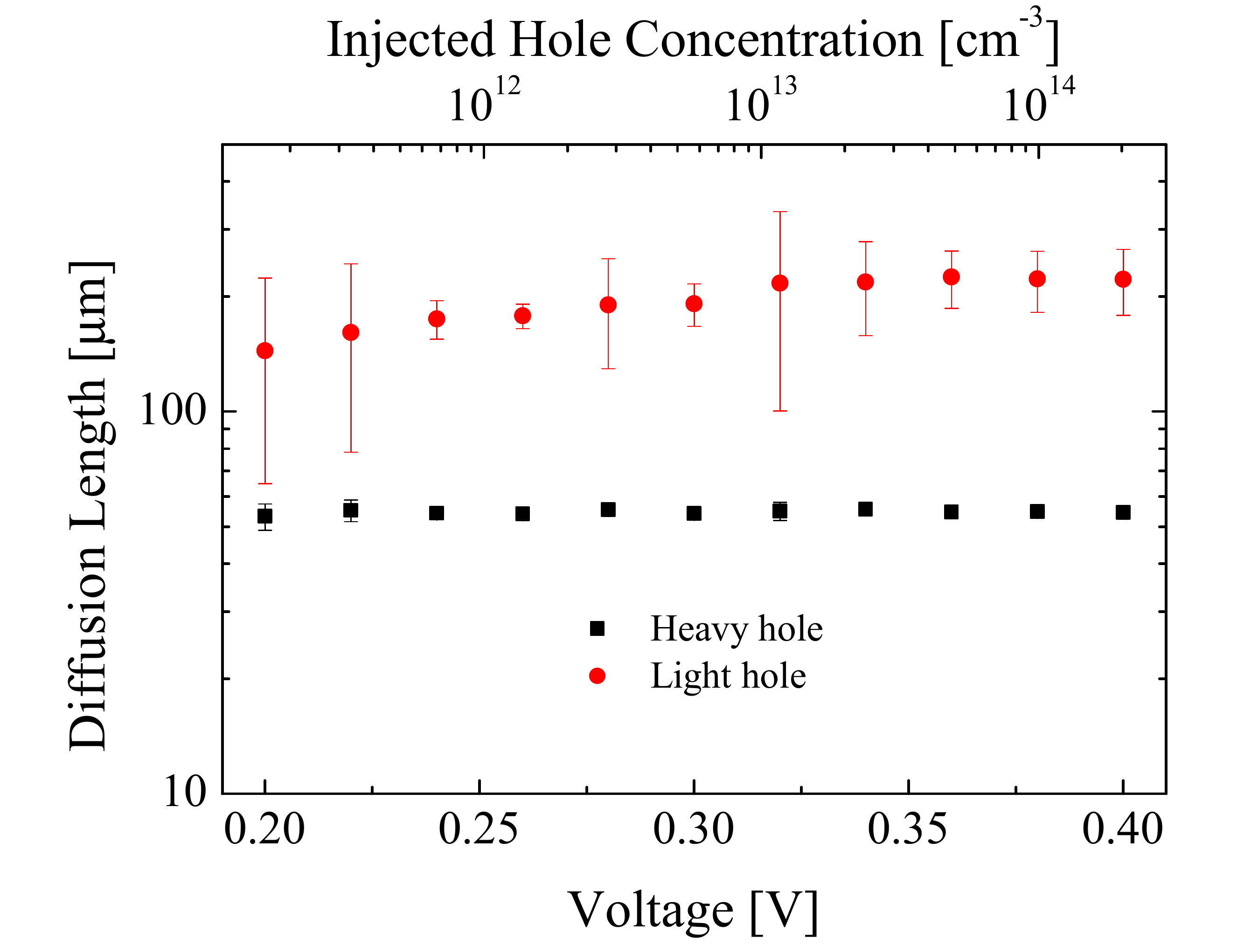}
    \end{center}
    \caption[example]
    {\label{fig:fig-4}
 Extracted hole diffusion length in n-InGaAs as a function of applied bias on a logarithmic scale (injected hole concentration shown on top axis). }
    \end{figure}
The fitting can be performed as a function of voltage to investigate the injection level dependence of the diffusion lengths for this sample.
 This also tests the ability of the method to extract consistent parameters.
 The results are illustrated in Fig.~\ref{fig:fig-4}, where the applied bias is shown on the bottom axis and the corresponding injected hole concentration is shown on the top axis.
 One can observe a near constant diffusion length across the voltage range of $0.2-0.4$ V for the heavy holes, which validates the robustness of the method.
 The light holes, however, show a trend of increasing diffusion length for increasing voltage, although it could be considered constant within the error bars.
 This may be due to the choice of the separation correction value, and/or by assuming that it is constant as a function of separation.
 It is further complicated by the random error in the current measurement.
 Improving the determination of the $W$ correction factor requires further investigation using careful device simulations.
 The standard deviation for heavy hole diffusion lengths ($<5\%$) is significantly smaller than for light holes (${\sim}25\%$). This is due to the heavy hole parameter fitting being based on sufficient data points, whereas light hole parameters rely on mainly the last two data points.
 More data points for larger interdiode separations would improve the accuracy of the extracted light hole diffusion lengths, and this may clarify the observed trend of increasing $L$ for increasing voltage.
 Lastly, the standard error is observed to vary considerably depending on the fit (even if it appears good by eye), and this is due to current measurement errors.
 Overall, the light hole diffusion lengths are ${\sim}3\times$ longer than the heavy hole diffusion lengths, which is a result of the larger light hole mobility (discussed next).
 Also, increasing the hole injection beyond $V>0.4$V leads to a breakdown of the model's low injection assumption.
 Lastly, fitting at sufficiently low voltage ($V<0.2$V) results in large uncertainties due to the significant scatter in the data arising from hysteresis in the the current measurement and the overall high noise in measuring currents below 10 pA.
 Improving the accuracy of the current measurement would result in more accurate diffusion length parameters.

  \begin{figure} [t]
   \begin{center}
    \includegraphics[width=0.48\textwidth]{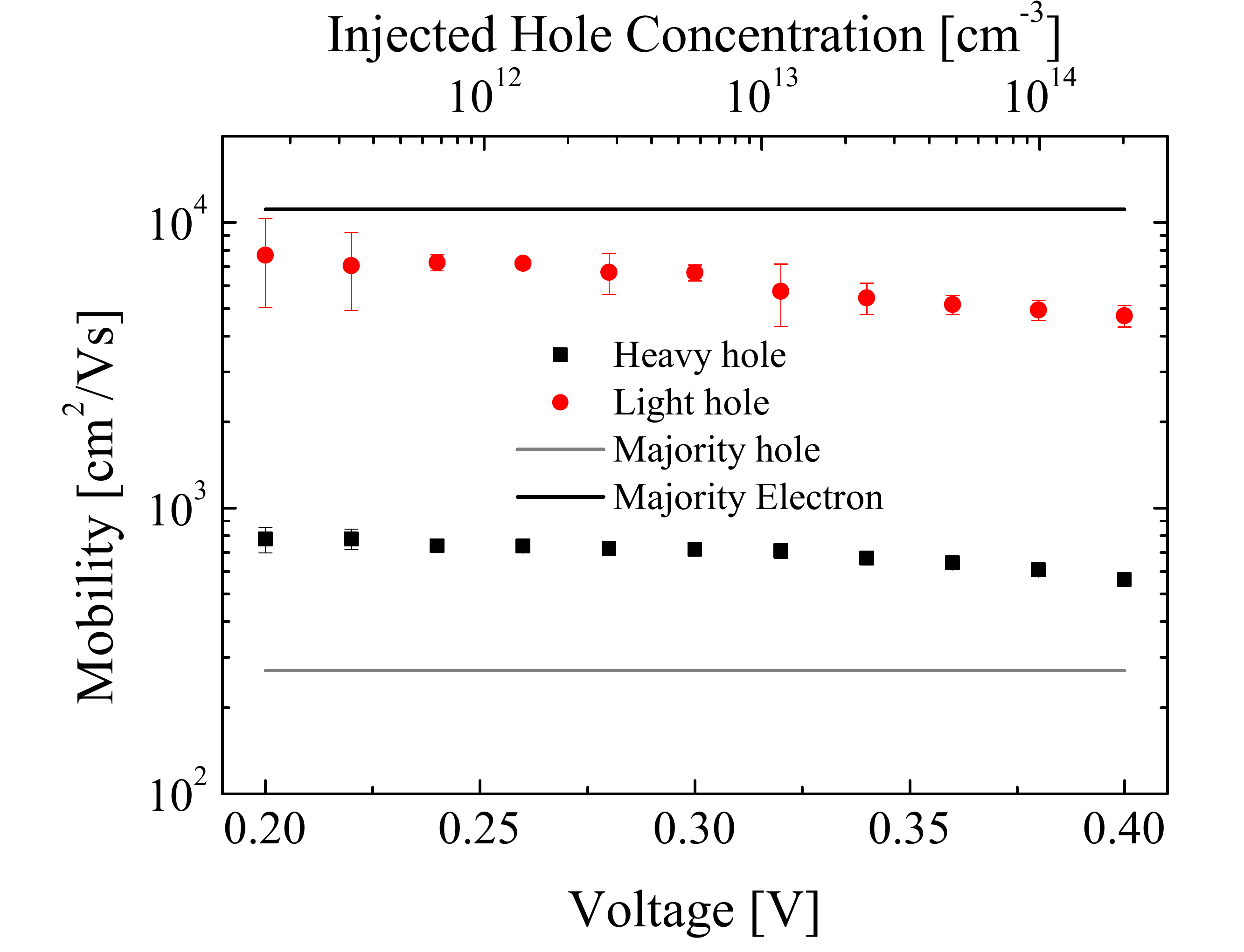}
    \includegraphics[width=0.48\textwidth]{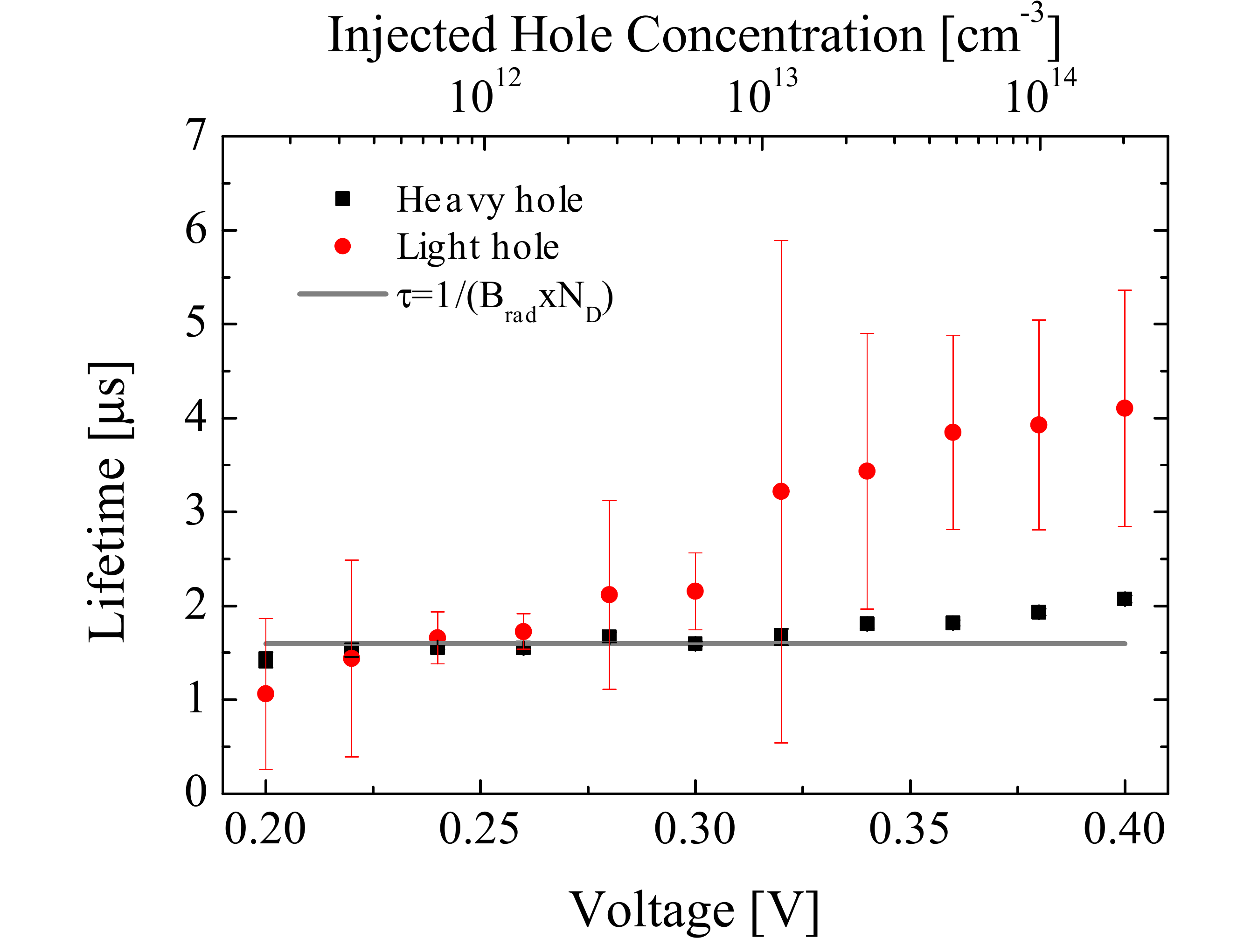}
    \end{center}
    \caption[example]
    { \label{fig:fig-5}
 Extracted a) mobility (on a logarithmic scale) and b) lifetime for heavy and light holes as a function of injection. Mobility plots includes the majority electron and hole Hall mobilities\cite{Sotoodeh} for the InGaAs doping of $N_D = 8.4 \times 10^{15} \text{cm}^{-3}$. The lifetime plot shows the expected radiative lifetime according to $\tau = \frac{1}{B_{rad}N_{D}}$, where $B_{\text{rad}}=0.75\times 10^{-10}\text{ cm}^{-3}$/s.}
    \end{figure}

The minority carrier mobilities are also obtained as a function of voltage from the fit to the data. The results are illustrated in Fig.~\ref{fig:fig-5}a. The average mobilities are $692\pm63$ and $6200\pm960\, \text{cm}^{2}\text{/Vs}$ for heavy and light holes respectively.
 The observed trends of decreasing mobility for increasing voltage were not expected.
 Further investigation into this is required to determine if this observation is real. Note that the mobility results depend strongly on the choice of band parameters (which dictate the calculated injection level) and temperature.
 The choice of separation correction $\delta$ primarily influences the light hole transport properties.
 Nevertheless, light holes are determined to be significantly more mobile than heavy holes by a factor of ${\sim}9$, which is expected based on their effective masses as well as their extracted diffusion lengths.
 The light hole mobility is somewhat smaller but on the same order of magnitude as the electron Hall mobility (${\sim}\,14000\, \text{ cm}^2/\text{Vs}$) for intrinsic material,\cite{Sotoodeh} considering their similar effective masses.
 Interestingly, the heavy hole mobilities are considerably larger than the majority hole Hall mobility of $269\, \text{ cm}^2/\text{Vs}$ for the corresponding sample doping level of $8.4\times 10^{15}\, \text{ cm}^{-3}$.\cite{Sotoodeh}
 Overall, the heavy and light hole mobilities fall within the range of hole and electron mobilities from Sotoodeh (see horizontal lines in Fig.~\ref{fig:fig-5}a).
 With respect to other published data, the reported hole mobility reported here is greater than the $425\, \text{cm}^{2}\text{/Vs}$ reported by Gallant and Zemel.\cite{Gallant}
 This emphasizes the importance of characterizing minority carrier mobilities for device design and simulation.
 Overall, careful temperature control is critical for this component of the study.
 Lastly, a better knowledge of band parameters would improve the extraction of the mobilities. 

Finally, Figure~\ref{fig:fig-5}b illustrates the extracted lifetimes for heavy and light holes as a function of injection. The heavy hole lifetime corresponds to the lifetime dictated by the radiative recombination coefficient $B_{\text{rad}} = 0.75\times10^{-10} \text{ cm}^3/\text{s}$ and the doping, indicated by the horizontal line. This is in agreement with other studies conducted on low-doped InGaAs.\cite{Wichman,Zielinski,Wintner,Walkerb} While technically this method cannot distinguish between bulk recombination in the layer and surface recombination at the hetero junction interfaces, the apparent dominance of radiative recombination justifies the assumption of negligible surface recombination. The lifetimes for heavy holes are more accurate than those for light holes. For the device reported here, as well as other devices, the light hole lifetime appears to be longer than for heavy holes. Again, more data in the range of the largest interdiode separations would be required to improve the accuracy of the lifetime for the light holes. An independent measurement of the effective lifetime (both heavy and light holes) could also reduce the fitting parameters, thereby potentially providing better results for the diffusion length and mobility.

In conclusion, a simple and nondestructive electrical method was proposed to extract minority carrier diffusion lengths, mobilities and lifetimes using long and thin diffused double-heterostructure diodes.
 The proposed method was demonstrated in low-doped n-InGaAs lattice matched to InP for a sample doped to $8.4\times 10^{15}\, \text{cm}^{-3}$.
 Heavy and light hole diffusion was observed as separate contributions, with a heavy hole diffusion length of $54.4\pm0.6\,$\textmu m (averaged over injection), and a light hole diffusion length of $195\pm26\,$\textmu m.
 The hole mobilities were extracted to be $692\pm63$ and $6200\pm960\, \text{cm}^{2}\text{/Vs}$ for heavy and light holes respectively. Ultimately, radiative recombination dominates the lifetime component, which is found to be $1.7\pm0.2\, \mu\text{s}$ for heavy holes and $2.6\pm1.0\, \mu\text{s}$ for light holes. The method is limited to low injection, and to diffusion lengths longer than the InGaAs layer thickness. 


\input{xyDiffusion.tex}

\end{document}

%% file: xyDiffusion.tex
\onecolumngrid
\newpage

\section{Appendix - not part of the APL article}
\pagestyle{append}

\Large
\noindent Assumptions used in Heavy and light hole minority carrier transport
 properties in low-doped n-InGaAs lattice matched to InP \\[0ex]
\normalsize
\begin{center}
Mike Denhoff, email: mwdenhoff@gmail.com \\
\today \\[2ex]
\end{center}

\twocolumngrid

\subsection{Introduction}

In this appendix, some of the assumptions used in the preceding APL article
 are explored.  This appendix is not part of the published article.
While the physical device on which the measurements were made is a
three dimensional (3d) structure,  the analysis is based on a one dimensional
(1d) analytical model.
The deviation of the 1d model from the 3d device is explored below.
The results will validate the analysis method used in the preceding article.

First a two dimensional (2d) finite element method model is introduced.
  While the article assumes
 there is no interface recombination, this effect can be included in the
 finite element model.
This 2d model can be used to simulate the full 3d device by doing it in
two steps, one in the $xz$ plane and one in the $xy$ plane.  
In this way, the one dimensional (1d) diffusion assumption is examined from two
 points of view.  One is the due to the finite thickness of the InGaAs layer
combined with the contacts being on the top surface of that layer.
The 1d model assumes that all the current is sourced or sinked precisely
at the edges of the contacts, while the 2d models shows 
 that the current source and collection occurs over a short length
from the edges of the contacts.  This edge effect can be expressed as
a ``transfer'' length which can be
estimated from finite element calculations.
Another thing that the 1d model assumes is that the line diodes are infinite
in length.  The actual devices have a finite length (750\,\textmu m) and
this will be studied using a finite element model ($xy$ plane) to calculate
the current that flows beyond the line diode ends.  Using this, small
corrections to the results of the 1d theory are found.

In addition to these geometry considerations, we had also assumed the low
 injection limit. This will be briefly discussed along with
what is  the range of currents where
the model is valid.

\subsection{Edge effects of top contacts}

There is an edge effect due to the current source and the current sink not
 strictly being at the edge of the contacts which are on the top of the
InGaAs layer.  Rather the current is spread over some distance beyond
these edges (similar to a transfer length).  Intuitively, one would think
this length would be less than the layer thickness.
An effective separation for the 1d model, $W$, can be defined as the actual
 separation between the edges of the two diodes, $S$, plus a correction, $\delta$,
so $W=S+\delta$.
  A finite element
model simulation is described next and  will be used to find an estimate
 of this extra length or length correction.

   \subsubsection{2d finite element model over thickness of InGaAs}

\begin{figure}[b]
\centering
\includegraphics[width=0.8\columnwidth]{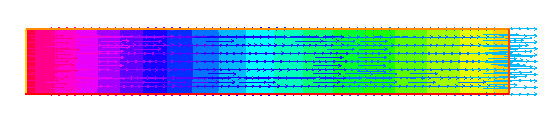}
\caption{Simple finite element model of hole current in a InGaAs layer
with contacts on either end, which is essentially the 1d case.
The shading represents the hole density decreasing from $p_{inj}$ at
the left edge to $p_0$ at the right edge.  The arrows show the current
which decreases from left to right due to recombination.
}%
\label{fig:hole-diff-recom-vel-02}
\end{figure}

The 2d model used here assumes that the line diodes are infinite in
length and will model a cross-section through the depth of the device.
In the low injection case, all current is due to the diffusion of minority
holes in the neutral lightly doped n-type InGaAs layer.
The motion of holes in the InGaAs layer is given by the diffusion equation
\begin{equation}\label{eq:xzdiff}
\frac{\partial^2p}{\partial x^2}+\frac{\partial^2p}{\partial z^2}
     + \frac{(p-p_0)}{L^2}=0,
\end{equation}
where $L^2=\tau D$.
$x$ is the distance along the layer and $z$ is the distance along the thickness
of the layer.  In order to solve this using the finite element method
method,\cite{reddy} the differential equation must be expressed in integral form
or the weak form, which is
\begin{eqnarray}\label{eq:xzweak}
\int_\Omega \left[ \frac{\partial w}{\partial x}\frac{\partial p}{\partial x}
    +  \frac{\partial w}{\partial z}\frac{\partial p}{\partial z}
    + \frac{wp}{L} - \frac{p_0}{L} \right]\partial x \partial y \nonumber \\
    -\int_\Gamma w\left[ n_x\frac{\partial p}{\partial x} + 
      n_z \frac{\partial p}{\partial z} \right]\partial \ell = 0.
\end{eqnarray}
$\Omega$ is the area on which the problem is defined, $\Gamma$ is
the boundary of that area and $w$ is a test function.
The boundary conditions must be set for the entire boundary by setting
either the value of $p$ (Dirichlet or essential boundary condition) or
the value of $\partial p / \partial n$ where $n$ is perpendicular to
the boundary (Neumann or natural boundary condition)
or a combination.
The Neumann condition is set by setting the value of $\partial p/\partial x$
and/or $\partial p/\partial z$ in the line integral in Eq.~\ref{eq:xzweak}.
This represents the hole current at that section of the boundary.

An example solution with the length of the layer
(ie. the separation, $S$) being 20\,\textmu m and
the thickness 2.7\,\textmu m is shown in Fig.~\ref{fig:hole-diff-recom-vel-02}.
This calculation was done using the Freefem++ finite element solving
 software.\cite{freefem}
The ends are the contacts with $p=p_{inj}$ on the left and $p=p_0$ on the right.
The current crossing the top and bottom boundaries is set to zero,
$\partial p/\partial z=0$.
The resulting current is uniform in the thickness but decreases from left to right
due to the hole recombination a rate set by $L$.

Surface recombination at the top and bottom interfaces can be simulated
using \cite{sze2}
\begin{equation}
qD\frac{\partial p}{\partial z} = qS_p(p_{surface}-p_0),
\end{equation}
where $S_p$ is the surface recombination velocity.
Then the line integral in Eq.~\ref{eq:xzweak} for the top boundary
becomes
\begin{equation}
-\int_{\Gamma_{top}} w\frac{S_p}{D}(p-p_0)\partial x.
\end{equation}
In this case some of the current stream lines bend into the top boundary
and $p$ is no longer uniform in the thickness but is lower at the boundary.
An example of Freefem++ code, ``hole-diff-recom-vel-02.edp'', is included
at the end of this appendix.
Solving this using both surface recombination and bulk recombination give results
that are consistent with using a single effective hole life time (and diffusion length).
Our experiment cannot separate the two effects.  It should be possible
to determine the bulk and surface recombination life times using samples
with different InGaAs layer thicknesses.

   \subsubsection{Edge effects of the contacts}

\begin{figure}
\centering
\includegraphics[width=1.0\columnwidth]{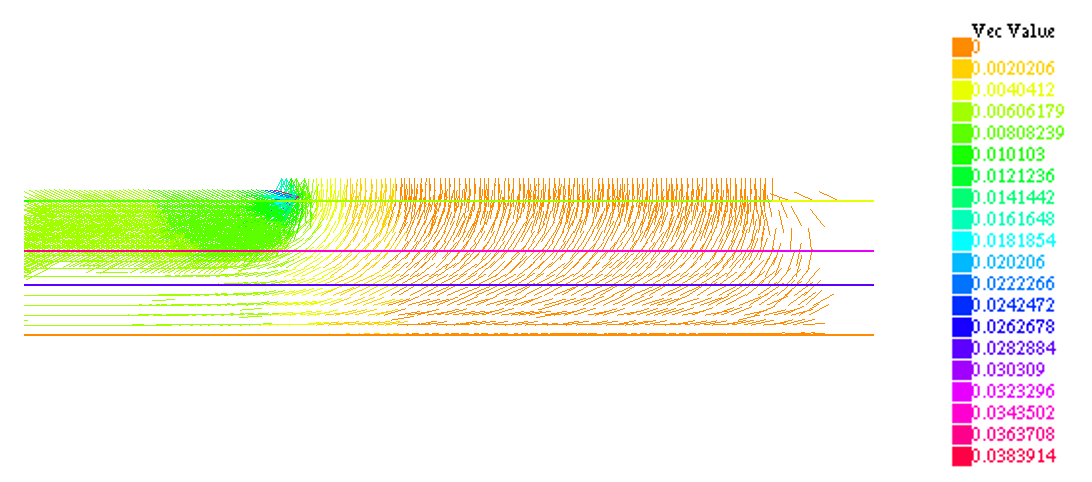}
\caption{Current crowding at the 10\,\textmu m long collecting electrode
  from a 2d, $x$-$z$, finite element simulation.
  The magnitude of the current density is color coded and is highest
 at the left edge which is closest to the injecting contact.
}%
\label{fig:hole-diff-06-20-curr}
\end{figure}

The assumption made in the analytic solution of the 1d diffusion equation
is that the entire current at the collecting electrode is collected at
the leading edge of the contact.  A finite element calculation
(hole-diff-07-50.edp) of the
hole current at a collecting contact on the top surface is given in 
Fig.~\ref{fig:hole-diff-06-20-curr}.
The largest current density is at the leading edge being about
$38$\,mA$\cdot$cm$^{-2}$.  Within 1\,\textmu m inside the edge
the current density has dropped more than an order of magnitude
 to $3$\,mA$\cdot$cm$^{-2}$
and after 2\,\textmu m it has dropped to close to zero.  This implies that
the position to be used as the boundary for the 1d problem is between 1
and 2\,\textmu m in from the edge of the contact.

\begin{figure}[t]
\centering
\raisebox{2.5ex}[0pt]{a)}\includegraphics[width=0.8\columnwidth]{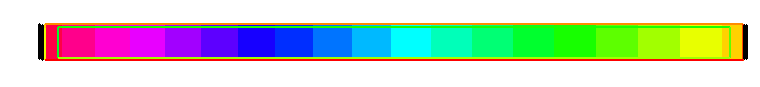}\\
\raisebox{2.5ex}[0pt]{b)}\includegraphics[width=0.8\columnwidth]{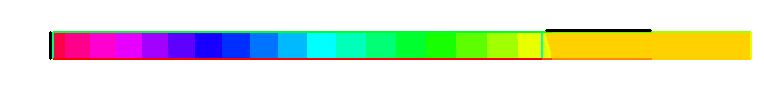}\\
\raisebox{2.0ex}[0pt]{c)}\includegraphics[width=0.8\columnwidth]{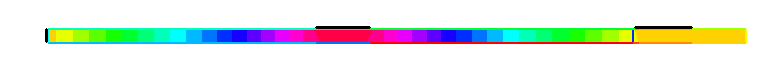}
\caption{a) Finite element solution for the hole concentration 
in the 1d case.  The injecting contact is on the left vertical edge and the
collecting contact is on the right edge.  The shaded contours show
 the hole concentration
varying uniformly from $p_{inj}$ on the left to $p_0$ on the right.
b) The black line on the top edge near the right end is now the collecting
contact.  The right edge of the contact is 50\,\textmu m from the injecting
contact on the left edge.  The shading shows the hole concentration is not
uniform over the thickness of the layer under the collecting contact.
c) The black line on the top in the middle is the injecting contact.  The InGaAs
layer has been extended to the left, with a second collecting contact at the
 left end, to create symmetry about the injecting contact.  The shading shows
that the hole concentration is highest under the injecting contact and
falls to the equilibrium concentration at the ends.
}%
\label{fig:hole-diff-05}
\end{figure}

The analytic 1d model for a single type of hole is given by
Eq.~\ref{eq:one}, which is repeated here
\begin{equation}
I = \frac{Aq(p_{inj}-p_0)D}{L\sinh(W/L)},
\label{eq:one-1d}\end{equation}
where $W$ is the distance between the injecting and collecting
contacts, $D$ is the diffusion constant, $L$ is the diffusion
length, and $A$ is the contact area.
The geometry of the finite element version of the 1d model is
shown in Fig.~\ref{fig:hole-diff-05}a
 (the same as Fig.~\ref{fig:hole-diff-recom-vel-02}).
The left end of the rectangle is the injecting contact and the right
end is the collecting contact.
The current at the right end calculated by Freefem++ is exactly the same
as the current found using Eq.~\ref{eq:one-1d}, which confirms the
finite element calculation.

In order to investigate the effect of one top contact, the collecting
electrode is added as a 10\,\textmu m segment on the top edge of the
finite element model, which is shown in Fig.~\ref{fig:hole-diff-05}b.
(Example Freefem++ code, ``hole-diff-07-50.edp'', is added at the
end of this appendix.)
With this geometry the finite element calculation is less accurate due
to current crowding and the change in flow direction
 at the leading edge of the collecting contact.
Here the hole concentration changes rapidly causing some numerical inaccuracy which
is magnified due to the current calculation using the derivative
of the concentration.
To investigate this, the current is found at a vertical line which is
1\,\textmu m in front of the edge of the collecting contact.
This should be very close to the collecting contact current, only
slightly higher since there will be a small amount of recombination
after the vertical line. 
For some value of $p_{inj}$, using an inter diode separation of
 $S=50$\,\textmu m in the finite element
model, the current calculated for this collecting contact is
 $I_c=6.476\times10^{-8}$\,A (at the top contact)
 or $I_v=6.492\times10^{-8}$\,A (at the vertical line).
The correct value is in between these.
This is slightly smaller than the current
in the 1d model for $W=50$, which is $I_{1d}=6.666\times10^{-8}$\,A.
However,  the 1d model for $W=51.2$\,\textmu m gives a current of
$I_{1d}=6.485\times10^{-8}$\,A.  This current matches the finite
element result for the top collecting contact, which implies a correction
to the length of 50\,\textmu m of 1.2\,\textmu m should be used in 1d model.
The same calculations for $S=150$\,\textmu m gives the same value
for the correction.  The correction is  only due to the distribution of the
hole concentration near the contact and not affected by the separation
between injection and collection (for $S\gg$ the layer thickness).

The next step is to add a more realistic injecting contact on the top surface
of the InGaAs layer as shown in Fig.~\ref{fig:hole-diff-05}c.
The separation between the right edge of the injecting contact and the
left edge of the collecting contact on the right hand side is 
50\,\textmu m.  The finite element calculated current at the collecting
contact is $I_c=6.304\times10^{-8}$\,A and $I_v=6.323\times10^{-8}$\,A
at the vertical line.
  This, again, is smaller than the
1d value for $S=50$ of $I_{1d}=6.666\times10^{-8}$\,A.
To match the finite element 2d result one needs to use $S=52.4$ which
gives $I_{1d}=6.312\times10^{-8}$\,A.
Then the correction for $S$ to account for both contacts is about
$\delta=2.4$\,\textmu m, which is twice the value that was found for
only the collecting contact in the above.

   \subsubsection{Discussion about the edge correction}

There are a couple of intuitive methods to check on the value of the
 end correction.

One is to assume that the best fit to the measured data will occur
when using the correct value for $\delta$.  We have found that the fitting
error given by the least squares routine is smallest for
 $\delta=2\pm 0.5$.  This is consistent with the estimate given above by
the finite element calculations.

The other method comes from considering that the relative error
in the value of the separation, $W$, is largest for the shortest
separation distance, ie.  in our case for $S=20$\,\textmu m.
One can fit the data using all the data points and then again using all
 the data points excluding the first point.  With the correct value of $\delta$,
the results of both fits should give the same values for $L_h$ and $L_l$
(see Eq.~\ref{eq:two-1d}).
The best value of $\delta$ can be found by trial and error.
The results varry somewhat for different data sets.
By this method we also find that $\delta=2\pm 0.5$, which is, again,
cosistent with the above.

\subsection{\label{sec:diff}Diffusion of holes in the \textit{x-y} plane}

\begin{figure}
\centering
\includegraphics[width=0.8\columnwidth]{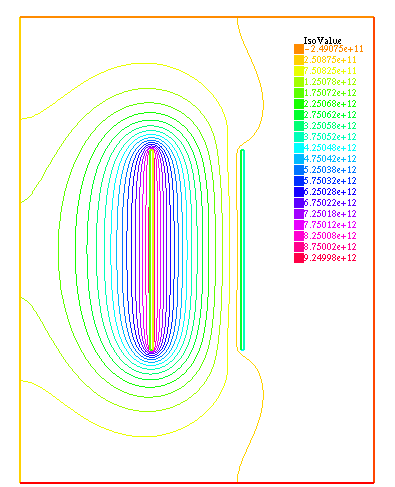}
\caption{Finite element solution for the hole concentration shown as
contour lines.  The separation between the 750\,\textmu m long
 diodes is 330\,\textmu m and the
diffusion length is $L=200$\,\textmu m.  The concentration is highest
 at the injecting line diode
on the left and lowest at the collecting diode at the right.
}%
\label{fig:hole-conc-05}
\end{figure}

Since the length of the line diodes is finite, the holes will diffuse
into the region beyond the diode ends.
One might expect that some current would be lost due to hole recombination
in this outer region.
This diffusion in the $xy$ plane can be modeled with a 2d finite element
model, assuming that the InGaAs layer is negligibly thin and that
all holes are emitted or collected exactly at the edges of the line diodes. 
Holes are injected by one line diode and diffuse away from it
 and are collected by a second line diode.  
 This implies that the
holes are confined to a thin layer in the $z$ direction and the
concentration is uniform in that direction.
The diffusion equation with recombination is
\begin{equation}
\frac{\partial^2p}{\partial x^2}+\frac{\partial^2p}{\partial y^2}
     + \frac{(p-p_0)}{L^2}=0,
\end{equation} 
where $L^2=\tau D$, is the diffusion length.
$p$ is the density of hole as a function of $x$ and $y$, $p_0$ is
the equilibrium hole density, 
$\tau$ is the hole life time, and D is the hole diffusion constant.

To solve this using the finite element method, the differential
 equation must be written in the weak (or integral) form,
\begin{equation}
\int_\Omega \left[ \frac{\partial w}{\partial x}\frac{\partial p}{\partial x}
    +  \frac{\partial w}{\partial y}\frac{\partial p}{\partial y}
    + \frac{wp}{L} - \frac{p_0}{L} \right]\partial x \partial y = 0.
\end{equation} 
The outer boundary of the area, $\Omega$, being simulated is assumed to
 have zero perpendicular current, so the line integral on the boundary is zero.
The two line diodes are defined by rectangular boundaries inside
 the large rectangle as shown in
 Fig.~\ref{fig:hole-conc-05}.
The boundary condition for the left hand, injecting, diode is the
injected hole density, $p=p_{inj}$, and for the right side, collecting
diode the boundary is set to the equilibrium hole density, $p=p_0$.
The solutions were calculated using the Freefem++ finite element
solving software.\cite{freefem}
An example Freefem++ input file, two-hole-diff-01v.edp, is included at the
end of this appendix.

\begin{figure}[t]
\centering
\includegraphics[width=\columnwidth]{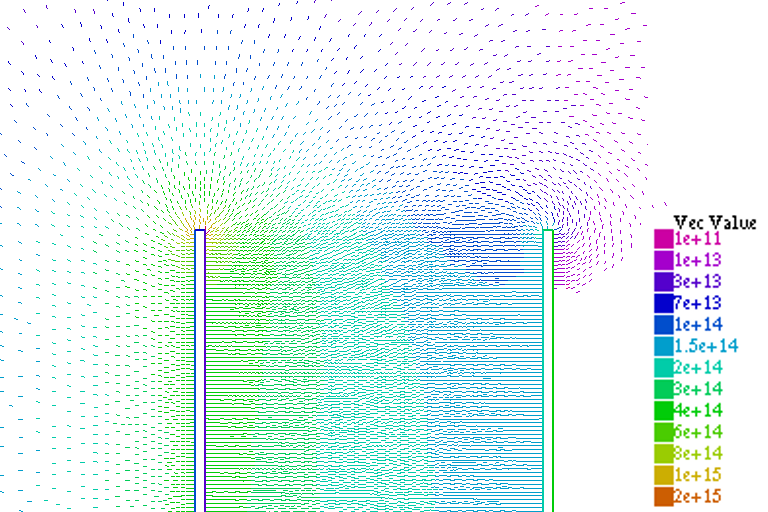}
\caption{Hole current stream lines.  Some current is collected at
 the top (and bottom) and on the ``back'' side of the collecting
diode.}%
\label{fig:hole-current-05}
\end{figure}

Current stream lines can be calculated from the derivatives of the
hole concentration.  These are drawn in Fig.~\ref{fig:hole-current-05}.
Current flows from all surfaces of the injecting diode, expands above
the region between the diodes so that  some current
flows to the top and bottom ends and even around to the back of the
 collecting diode.  This results in collected current being slightly
larger than that predicted by the pure 1d formula.  This is true
except for small values of $L$ relative to $S$,
 where a large fraction of holes are lost
to recombination in the area outside the ends of the line diodes.

   \subsubsection{Numerical comparison of 2d and 1d current results}

Collected current can be calculated for both the 2d and 1d methods in order to
estimate the error made by using the 1d model due to end effects (which are included
in the 2d finite element calculation).  For these calculations the outer boundary
of the finite element area was doubled compared to that shown in
Fig.~\ref{fig:hole-conc-05} such that the boundary is more than a few
diffusion lengths away from the line diodes. 
 Some results are presented in Table~\ref{tab:2d1d}.
For separations of 20 and 100\,\textmu m, the 1d model
 underestimates the actual current as calculated by the 2d model.
  The underestimation is larger as the diffusion length
increases.  This is due to the holes being able to diffuse further in the region above
(and below) the ends of the line diodes.
The case for $S=330$\,\textmu m and $L=60$\,\textmu m, the opposite occurs, that is 
the 1d model gives a larger current than the 2d model.  This is due to more of the
holes that diffuse into the regions above and below the ends of the line diodes
being lost to recombination before they reach the collecting diode.
In general, the relative error varies from less than 1\% to as much as
7\%.   The effect this has on analyzing the experimental results in investigated
in the next section.

\begin{table}[b]
\caption{2d (finite element, with line diode length of 750\,\textmu m)
 and 1d (Eq. (\ref{eq:two-1d})) calculations of current for various separations
    and diffusion lengths}\label{tab:2d1d}
\begin{tabular}{c||c|ccc}
Separation & Model & $L=60$\,\textmu m & $L=200$\,\textmu m & $L=300$\,\textmu m \\
\hline \hline
                                    & $I_{2d}$ &\rule{0pt}{2.5ex} $8.670\times10^{-9}$
                                        & $8.896\times10^{-9}$ & $8.933\times10^{-9}$  \\
 \cline{3-5}
\raisebox{1.5ex}[0pt]{20\,\textmu m} & $I_{1d}$ &\rule[-0ex]{0pt}{2.5ex} $8.550\times10^{-9}$
                                        & $8.694\times10^{-9}$ & $8.702\times10^{-9}$ \\
\hline 
                                     & $I_{2d}$ &\rule{0pt}{2.5ex} $1.141\times10^{-9}$
                                        & $1.751\times10^{-9}$ & $1.820\times10^{-9}$ \\
 \cline{3-5}
\raisebox{1.5ex}[0pt]{100\,\textmu m}& $I_{1d}$ &\rule{0pt}{2.5ex} $1.137\times10^{-9}$
                                        & $1.671\times10^{-9}$ & $1.710\times10^{-9}$ \\
\hline
                                     & $I_{2d}$ &\rule{0pt}{2.5ex} $2.228\times10^{-11}$
                                        & $3.477\times10^{-10}$ & $4.547\times10^{-10}$ \\
 \cline{3-5}
\raisebox{1.5ex}[0pt]{330\,\textmu m}& $I_{1d}$ &\rule{0pt}{2.5ex} $2.373\times10^{-11}$
                                        & $3.473\times10^{-10}$ & $4.347\times10^{-10}$ \\
\hline
\end{tabular}
\end{table}

\begin{figure}[b]
\centering
\includegraphics[width=\columnwidth]{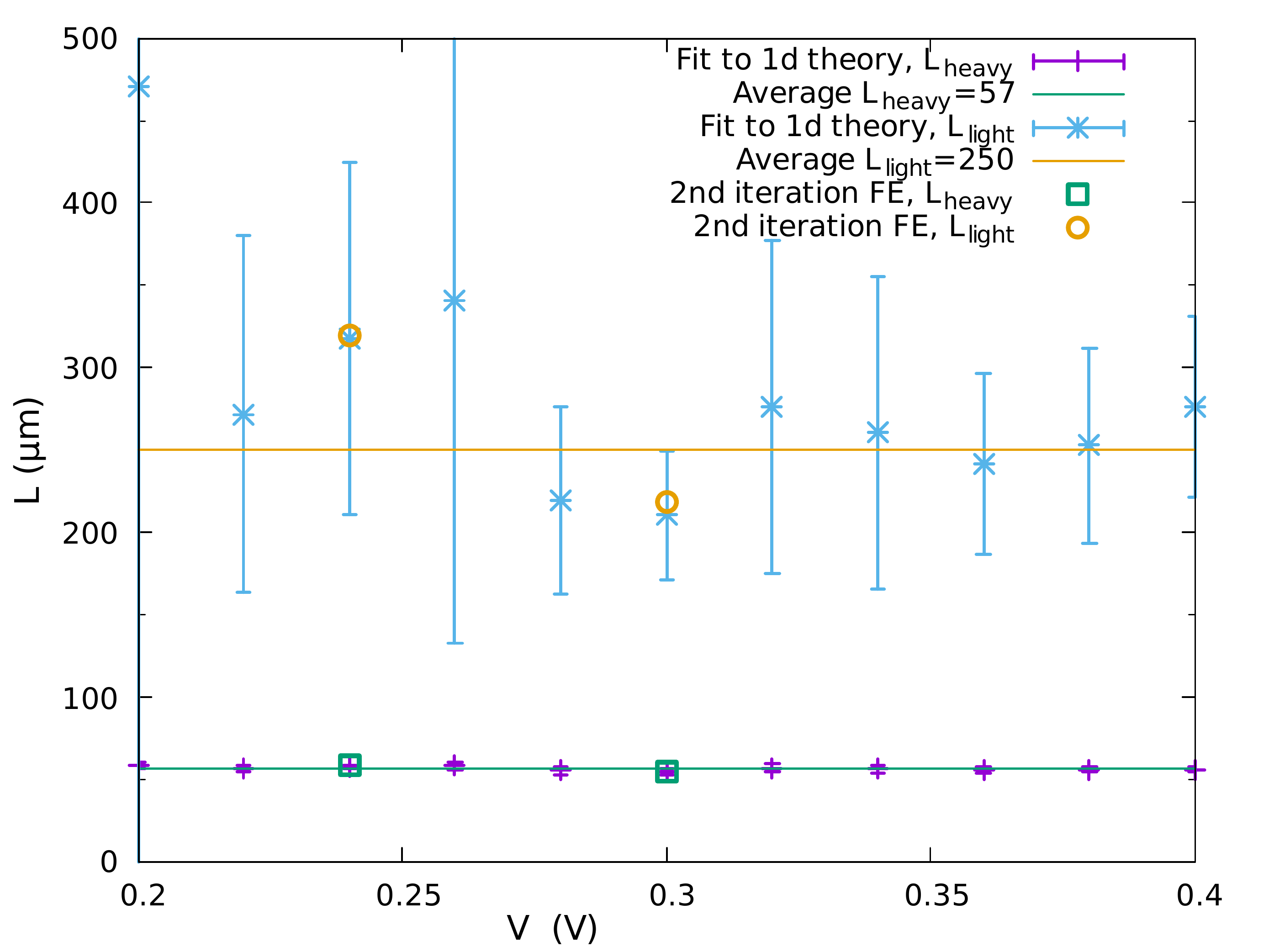}
\caption{Diffusion lengths extracted from fits of the 1d equation
to measured currents on sample 31.  In addition, improved values
found from a second itereation using the 2d model for $V=0.24$ and $0.30$ are plotted.
}%
\label{two-31-S+2-Lh-Ll}
\end{figure}

   \subsubsection{Comparison with current measurements}

Since the 2d calculation is more realistic (and so more accurate)
 than the 1d theory, ideally,
 one would fit the measured data the the 2d model.  This would have to 
be done by trial and error guessing test values for the four fitting values
and doing many repeated finite element calculations.  This would be onerous
compared to using the analytic 1d theory with which can use available
non-linear least square fitting routines.  In this section, fits to the 1d
 and 2d models will be compared and an improved result from the 2d model
is obtained.

The finite element and 1d results will be compared using values found
in the analysis of experimentally measured results.
To do this, the finite element solution can be found separately for heavy and light
holes with the collected current being the sum of both contributions.
An example Freefem++ input file, two-hole-diff-01v.edp, is included at the
end of this appendix.
This can be compared to results of the 1d solution which is
\begin{equation}
I = \frac{Aq(p_{h\cdot inj}-p_{h0})D_h}{L_h\sinh(S/L_h)}
   + \frac{Aq(p_{l\cdot inj}-p_{h0})D_l}{L_l\sinh(S/L_l)},
\label{eq:two-1d}\end{equation}
where the $h$ and the $l$ subscripts refer to heavy or light hole
values.

\begin{figure}
\centering
\includegraphics[width=\columnwidth]{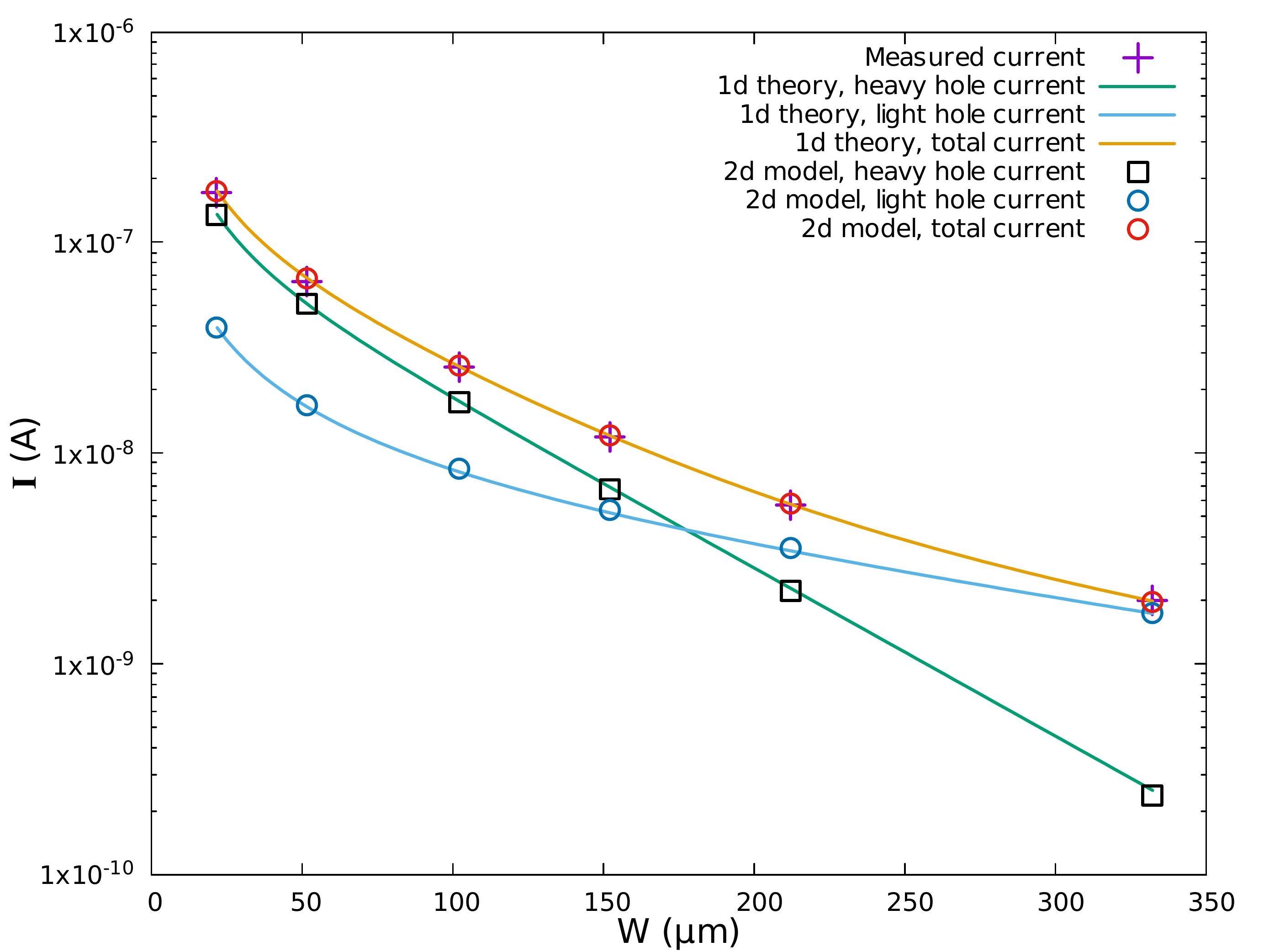}
\caption{Measured current (\textcolor[rgb]{0.5,0,0.5}{+}) finite element current
 (\textcolor{red}{$\circ$}) for $V=0.3$\,V.  The solid lines are the heavy hole,
light hole, and total current for the 1d theory fitted to the currents
calculated by the finite element method.
}%
\label{two-IvsS+2-31-Freefem-L-D-V_3-ni6_1386-T296}
\end{figure}

Results for a device from a different wafer (with an InGaAs layer doping
level of $9.4\times10^{15}$) than the results presented
in the main APL paper, analyzed in the same way using the 1d theory with
 $\delta = 2$\,\textmu m are plotted in Fig.~\ref{two-31-S+2-Lh-Ll},
which shows similar results as those in Fig.~4 of the APL article.

\begin{figure}[b]
\centering
\includegraphics[width=\columnwidth]{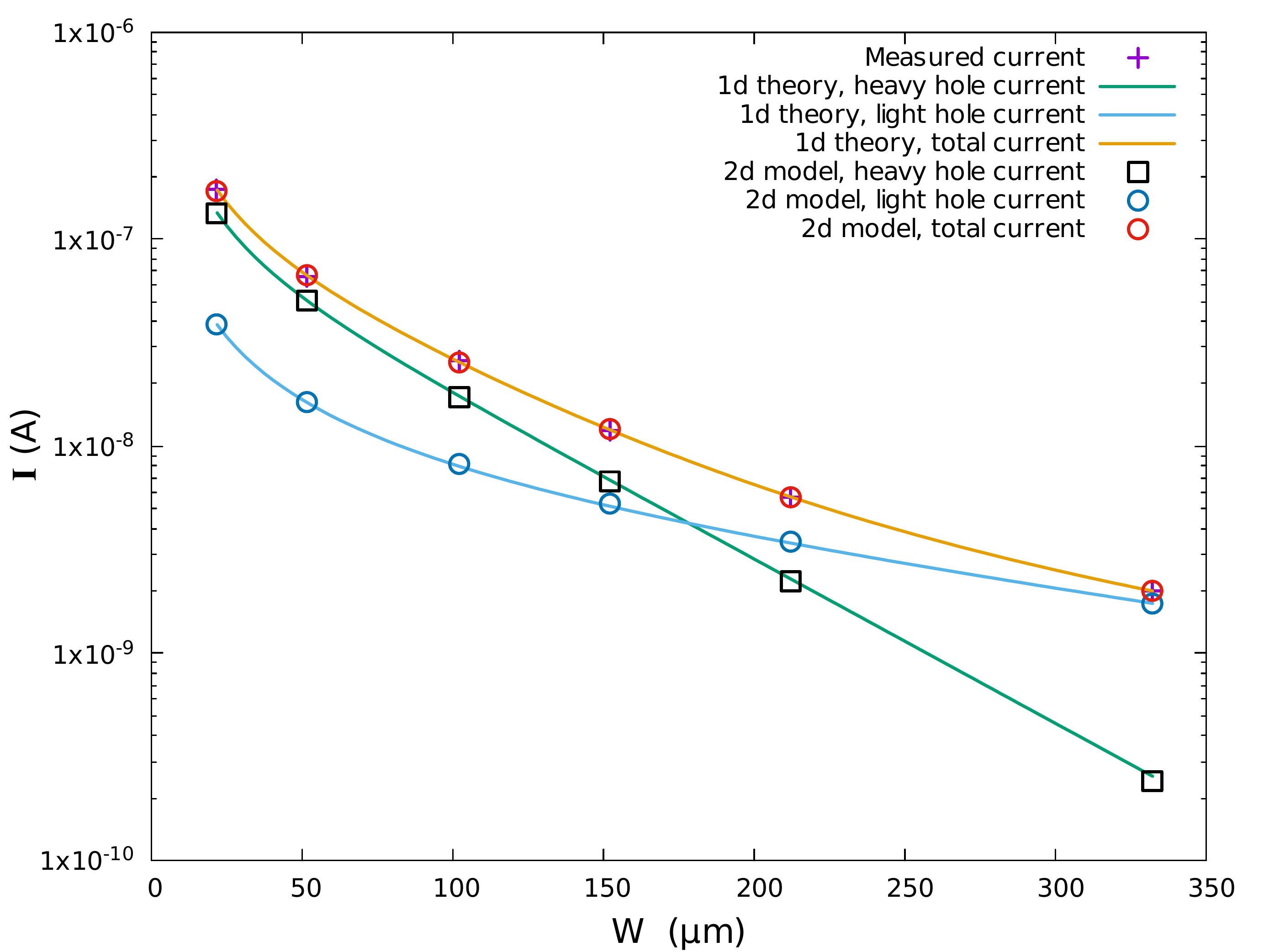}
\caption{Measured current (\textcolor[rgb]{0.5,0,0.5}{+})and the second iteration
 finite element current
 (\textcolor{red}{$\circ$}) for $V=0.3$\,V.  The solid lines are the heavy hole,
light hole, and total current for the 1d theory fitted to the currents
calculated by the finite element method.
}%
\label{two-IvsS+2-31-Freefem2-L-D-V_3-ni6_1386-T296}
\end{figure}

As an example, calculations are done for the case of an applied
voltage of 0.3\,V. 
 The fit of the 1d theory, Eq.~(\ref{eq:two-1d}), to experiment gave $L_h=54.66$,
$L_l=210.7$, $D_h=17.35$, and $D_l=149.1$.
 (Plotted in Fig.~\ref{two-31-S+2-Lh-Ll} for $V=0.3$.
These values are used in the finite element model to calculate
the collected current for various values of separation.
The outer boundaries in the finite element problem were expanded
 to be twice the size as that shown in Fig.~\ref{fig:hole-conc-05}
(1000\,\textmu m beyond the ends of the line diodes) in
order for the boundary to be more than 3 diffusion lengths from the
line diodes.
The measured and the 2d calculated currents are in good agreement on
the scale of the plot in
Fig.~\ref{two-IvsS+2-31-Freefem-L-D-V_3-ni6_1386-T296}.
  However, looking at the actual numbers, the 2d calculated
 currents are all slightly larger
 than the measured values (on the order of 1\%), with the values at
 $W=332$\,\textmu m being
 closest to each other.  The 2d model with the above values of $L$ and
$D$ is not an optimal fit to the measured data.
Then the calculated 2d current was fit to the 1d equation.
This results in values  $L_h=54.33$\,\textmu m $L_l=202.9$\,\textmu m,
 $D_h=17.64$\,\textmu m and $D_l=152.9$\,\textmu m.
These are slightly different than the input values 
 used to calculate the 2d current.
The new $L_h$ and $L_l$ are 0.6\% and 3.8\% smaller and  $D_h$ and $D_l$ are
1.6\% and 2.5\% larger, respectively.
The 1d theory is plotted with these values in
 Fig.~\ref{two-IvsS+2-31-Freefem-L-D-V_3-ni6_1386-T296}.
Also plotted in Fig.~\ref{two-IvsS+2-31-Freefem-L-D-V_3-ni6_1386-T296}
 are the 2d calculated heavy and
light hole currents compared to the 1d fit values of these.
They are in good but not exact agreement, showing that the 1d model 
is a reasonable approximation to the 2d calculations.

In order to achieve an improved  fit of the 2d FE model to the measured data,
a second iteration of the finite element calculation was done.
The above two sets of $L$ and $D$ values were used as a guide to give
 improved values to use in the finite
element calculation.  These improved input values are
$L_h=54.99$\,\textmu m, $L_l=218.5$\,\textmu m,
 $D_h=17.06$\,\textmu m, and $D_l=145.3$\,\textmu m.
The new 2d current calculated values are plotted in
Fig.~\ref{two-IvsS+2-31-Freefem2-L-D-V_3-ni6_1386-T296}.
While this looks identical to Fig.~\ref{two-IvsS+2-31-Freefem-L-D-V_3-ni6_1386-T296},
it is reproduced here to demonstrate how close the two results are.
Again the total finite element currents match the measured values
on the scale the plot.  Looking at the actual numbers,
the 2d calculation is now a very good fit to the data. 
The quality of this fit is similar to the original fit of the 1d
theory to the data, so the fitting uncertainties would imply that
the 2d and 1d models are equally good.
If, due to the physics, one considers the 2d FE model to be more realistic
 than the 1d theory,
then the new values for $L$ and $D$ used in the second finite element
calculation are improved values for the actual sample.
Also plotted are the 1d currents from the fit of Eq.~\ref{eq:two-1d}
 to the new finite element calculated values.
These new 1d fit values are similar to the original values from the fit of the
1d theory to the measured currents, but they have smaller fitting
errors.  This fitting error is due to the difference between the 2d
 model and the 1d model. It does not represent the experimental
measurement errors.

The above improved result plus the similarly calculated improved result
for $V=0.24$ are plotted in Fig.~\ref{two-31-S+2-Lh-Ll}.
 The difference between the 2d analysis and the 1d analysis is small,
 well within the
uncertainties for this experiment, so using the 1d theory to
analyze the data is satisfactory.

As was shown above corrections to the 1d theory due both to the
transfer length at the edge of the contacts ($xz$) and diffusion
 in the $xy$ plane at the line
ends were small, about 1\%.
However, since the fit is using 4 parameters, the fit results are
sensitive to these small corrections.  The experimental errors
in our measurements mask these small corrections, but with a
more accurate measurement of the currents, the 2d corrections may
be more important.

   \subsubsection{Note on the fitting method}

In this appendix, the fits of the 1d theory to the current measurements
were done using the nonlinear least-squares Marquardt-Levenberg
 algorithm in gnuplot.~\cite{gnuplot}
In order to give similar weight to all data points, the logarithm
of the measured currents was fit to the logarithm of Eq.~\ref{eq:two-1d}.
Due to the relatively large number (4) of fitting parameters and the fact
that the logarithm of current plotted against $W$ does not deviate too 
much from a straight line, the fitting is difficult.
The result is that small experimental errors in the current points
leads to large uncertainty in the resulting fit.
In a few cases, the fitting algorithm could not find a stable result.
This means that the fits had to be monitored manually to insure
reasonable results.

Compounding the above is that the 1d theory does not exactly represent
the actual 3d device (as was shown in previous sections).
This means there will be some fitting error even in the ideal case
of no experimental current measurement errors.

\subsection{Low injection assumption}

As well as the 1d assumption, we also use the assumption of low injection.
In this case the minority hole concentration must be much less than the
 electron concentration.
In the case of medium injection, the injected hole density will be somewhat
lower than the prediction of the formula used for low injection.\cite{cristea}
In our case the decrease in injected hole concentration would be about 1\%
at a junction voltage of 0.37\,V.
Low injection also allows us to assume that the electric field is zero
in the neutral region, so that the current
is only due to diffusion.  It also implies that the electron concentration
is unperturbed so that electrons do not contribute to the current.
The medium injection current is discussed in some detail by Kasap and Tannous,\cite{kasap}
who give some ways to calculate when these effects become important.

A second consequence of the move away from small injection and the existence
of an electric field in the neutral region is that the whole of the applied
voltage is no longer across the diode junction.
In our case, the current from the injecting diode to the ground must cross
the bottom heterojunction and there could be a potential drop there.
In addition other sources of series resistance, such as contact
resistance, will also lower the junction voltage as current gets
larger.
This affects the calculation of the hole concentration under the injecting
junction.  Above low injection, our calculation of $p_{inj}$ will be too
high.  This correction of the diode potential will lead to a lower $p_{inj}$,
which does not affect the value of $L$, but it does affect the
determined values of $D$ (or mobility) and $\tau$.
Since our device structure is quite complicated, an attempt to calculate 
the voltage correction will not be done.
However, in our experiments, the $IV$ curves drop below the ideal exponential
dependence by about 1\% at about $V=0.3$\,V,  indicating the beginning
of the aforementioned effects.

   \subsection{Band parameters}

Values for the electron and hole effective masses and the band gap
 have been assumed.
These values are used in the calculation of $p_{h\cdot inj}$ and
$p_{l\cdot inj}$.
This directly effects the values of $D$ (or $\mu$) and $\tau$, but it does not
effect the value of $L$.
This can be seen by writing Eq.~\ref{eq:two-1d} as
$I=C_h/\sinh(S/L_h)+C_l/\sinh(S/L_l)$, where $C_h$ and $C_l$ are constants.
The effective masses are not well known, with values found in the
 literature varying by 10 to 20\%.
This means the determination of $D$ (or $\mu$) and $\tau$ have this same
uncertainty. 

Another point is that  $p_{h\cdot inj}$ and
$p_{l\cdot inj}$ are strongly dependent on temperature, $T$, so
that the sample temperature must be accurately measured.\\[1ex]

   \subsection{Future}

Finally, a few possible extensions of the above experiment will be
mentioned, without any attempt to work out the details.

Another way to improve the experiment would be to measure some of the
 parameters independently.  For example, if experiment or theory could
fix the ratio of the life times (in the simplest case, $\tau_h=\tau_l$),
then there would be only three independent fitting parameters.
This greatly reduces the rms fitting errors.
It may also be possible to measure the life times on the same devices
using time dependent measurements.  The $\tau_h$ and $\tau_l$ would be
known and the only fitting parameters would be $L_h$ and $L_l$, again leading
to better fitting results.

A mesa etch of the whole set of line diode devices would stop current spreading
beyond the ends of the line diodes in the $xy$ plane.
A simple analysis would require that there was no recombination on the etched sidewalls.

\section*{Acknowledgment}

The electrical measurements were done by Alexandre Walker.  The method
of analysis was jointly developed by Alexandre Walker and Mike Denhoff.

\onecolumngrid

  \subsection{Example code}

The following examples of Freefem++ input code were developed until
they barely worked.  They have not been carefully tested or optimized for
correctness or accuracy.

   \subsubsection{hole-diff-recom-vel-02.edp}

The following is an example input file to Freefem++ which was used
to calculate Fig.~\ref{fig:hole-diff-recom-vel-02} (with Sp = 0.0).

\lstset{language=C++,columns=fullflexible,showstringspaces=false,keywordstyle={}}
\begin{lstlisting}
// hole-diff-recom-vel-02.edp, Copyright Michael Wayne Denhoff 2017.
// hole diffusion in cross-section of InGaAs with surface recombination
// on the top and bottom surfaces.   

real L = 20.0e-4; // diffusion length in cm, L^2=tD
real D = 10.0000; // diffusion const
real Sp = 1e2; // surface recombination velocity, cm/s
real p0 = 8.7446e8; // equilib. hole conc
real p1 = 9.5834e12; // hole conc at injecting contact
real S = 20.0e-4; // separation of the line diodes, cm
real T = 1.35e-4; // half the InGaAs thickness, 2.7e-4/2, cm
//real W = 0.3e-4; // depletion width under injecting diode, cm

border G1(t=0,S) {x=t;y=-T; }
border G2(t=-T,T) {x=S;y=t; }
border G3(t=S,0) {x=t;y=T; }
border G4(t=T,-T) {x=0;y=t; }

// G4 is the injecting contact
// G2 is the collecting contact

mesh Th = buildmesh(G1(60)+G2(10)+G3(60)+G4(10));
fespace Vh(Th,P1);
 Vh p, w;

plot (Th, wait=true, value=true);

problem diffp(p,w,solver=LU) =
       int2d(Th)( dx(w)*dx(p) + dy(w)*dy(p))
       + int2d(Th)(w*p/L^2)
       - int2d(Th)(w*p0/L^2)
     -int1d(Th,G1)(-w*Sp/D*p)
     -int1d(Th,G1)(w*Sp/D*p0)
     -int1d(Th,G3)(-w*Sp*p/D)
     -int1d(Th,G3)(w*Sp*p0/D)
//       + on(G1,p=p0) + on(G3,p=p0)
       + on(G2,p=p0) +on(G4,p=p1);

diffp;

plot (Th, p, value=true, coef=1, wait=true);

cout <<"holes"<<p(0,0)<<"   "<<p(0,S)<< endl;

// calculate the hole current at the collecting contact, for 7.5e-2 cm long diode
real Ih = int1d(Th,G2)(-1.6e-19*D*dx(p) );
cout <<"end current  "<<",  Ih  "<<Ih<<", Ih*7.5e-2  "<<Ih*7.5e-2<< endl;

// x and y hole current densities
Vh jpx = -1.6e-19*D*dx(p);
Vh jpy = -1.6e-19*D*dy(p);
plot(p, [jpx,jpy], coef=8e-1, wait=1, value=true);
//plot( jpx,  wait=1, value=true);
cout << "J(0) = "<<jpx(0,0)<<",  J(S) = "<<jpx(S,0)<< endl;

/*
{
ofstream file("conc-sv.dat");
file << "# hole concentraton along x-axis, between contacts." << "\n";
file << "# x um     p" << "\n";
for (int i=0;i<=50;i++){
file << i*(S*1e4/50) << "     " << p(i*(S/50),0) << "\n";
    }
};
*/
\end{lstlisting}

    \subsubsection{hole-diff-07-50.edp}
The following Freefem++ file was used to generate Fig.\ref{fig:hole-diff-05}b
and calculate the current for this case.

\begin{lstlisting}[showstringspaces=true]
// hole-diff-07-50.edp, Copyright Michael Wayne Denhoff, Sept 2017
// end effect of collecting diode which is on top surface

real L = 71.69e-4; // diffusion length in cm, L^2=tD
real D = 11.6277; // diffusion const  11.5547
real tau = 4.42e-6; // hole life time
real p0 = 8.7446e8; // equilib. hole conc
real p1 = 9.5834e12; // hole conc at injecting contact
real T = 1.35e-4; // half the InGaAs thickness, 2.7e-4/2, cm
real S = 49.0e-4; // separation of the line diodes minus 1e-4, cm
real U = 1.00e-4; // segment before start of collecting contact
real V = 1.0e-3; //  width of diode contact, cm
real W = 1.0e-3; // length beyond contact

border G1(t=0,S) {x=t;y=-T; }
border G2(t=S,S+U) {x=t;y=-T; }
border G3(t=S+U,S+U+V) {x=t;y=-T; }
border G4(t=S+U+V,S+U+V+W) {x=t;y=-T; }
border G5(t=-T,T) {x=S+U+V+W;y=t; }
border G6(t=S+U+V+W,S+U+V) {x=t;y=T; }
border G7(t=S+U+V,S+U) {x=t;y=T; }
border G8(t=S+U,S) {x=t;y=T; }
border G9(t=S,0) {x=t;y=T; }
border G10(t=T,-T) {x=0;y=t; }
border H1(t=-T,T) {x=S;y=t; }

// G10 defines the injecting contact
// G7 defines the collecting contact
/* H1 defines a vertical line 1 micron before the
         leading edge of the collecting contact */

mesh Th = buildmesh(G1(100)+G2(6)+G3(100)+G4(40)+G5(50)
        +G6(40)+G7(100)+G8(4)+G9(100)+G10(50)+H1(50));

fespace Vh(Th,P1);
 Vh p, w;

plot (Th, wait=true, value=true);

problem diffp(p,w,solver=LU) =
       int2d(Th)( dx(w)*dx(p) + dy(w)*dy(p))
       + int2d(Th)(w*p/L^2)
       - int2d(Th)(w*p0/L^2)
       + on(G7,p=p0) + on(G10,p=p1);

diffp;

// use adaptmesh to refine the mesh to get better accuracy
real error = 0.01; // chosen by trial and error
real NBVX = 50000; // chosen by trial and error
for (int i=0;i<13;i++)
{
Th=adaptmesh(Th,p,err=error,nbvx=NBVX);
diffp;
error=error/2;
};
plot (Th, p, value=true, coef=1, wait=true);

cout <<"holes  "<<p(0,0)<<"   "<<p(S+U,0)<<"   "<<p(S+U,T)<< endl;

// calculate the hole current at the collecting contact
real Ihc = int1d(Th,G7)(-D*dy(p))*1.6e-19*7.5e-2; // length of line diode= 7.5e-2
cout <<"Collecting contact current  "<<Ihc<< endl;

// calculate the hole current at the vertical before the collecting contact
real Ihv = int1d(Th,H1)(-D*dx(p))*1.6e-19*7.5e-2; // length of line diode= 7.5e-2
cout <<"Before Collecting contact current  "<<Ihv<< endl;

// calculate the hole current at the vertical injecting contact
real Ihi = int1d(Th,G10)(-D*dx(p))*1.6e-19*7.5e-2; // length of line diode= 7.5e-2
cout <<"Injecting contact current  "<<Ihi<< endl;

// x and y hole currents
Vh jpx = dx(p);
Vh jpy = dy(p);
plot(p, [jpx,jpy], coef=2e-22, wait=1, value=true); 
//plot( jpx,  wait=1, value=true);

\end{lstlisting}

   \subsubsection{two-hole-diff-01v.edp}
This Freefem++ file was used to calculate the results in
Fig.~\ref{two-IvsS+2-31-Freefem-L-D-V_3-ni6_1386-T296}.

\begin{lstlisting}
// two-hole-diff-01v.edp, Copyright Michael Wayne Denhoff Sept 2017
// diffusion in x-y plane of heavy and light holes
// diode length=750 micron,
//   n-type doping=6.64e15->p0=4.2191e8,p1=5.3763e12, V=0.3

real L = 54.66e-4; // diffusion length in cm, L^2=tD, heavy holes
real D = 17.3500; // diffusion const, heavy holes
real p0 = 4.21915e7; // equilib. heavy hole conc
real p1 = 5.3762e12; // heavy hole conc at injecting contact
real Ll = 210.7e-4; // diffusion length in cm, L^2=tD, light holes
real Dl = 149.1000; // diffusion const, light holes
real pl0 = p0/30.619; // equilib. light hole conc
real pl1 = p1/30.619;// light hole conc at injecting contact
real S = 52.0e-4; // separation of the line diodes, cm
real W = 7.50e-2; // length of the diodes, cm.
real V = 10.00e-2; // extended space beyond the diodes.
real Y = 1.0e-3; // width of line diodes

border G1(t=-V,S+V) {x=t;y=-V-W/2; }
border G2(t=-V-W/2,V+W/2) {x=S+V;y=t; }
border G3(t=S+V,-V) {x=t;y=V+W/2; }
border G4(t=V+W/2,-V-W/2) {x=-V;y=t; }
border H1(t=-W/2,W/2) {x=0;y=t; }
border H2(t=0,-Y) {x=t;y=W/2; }
border H3(t=W/2,-W/2) {x=-Y;y=t; }
border H4(t=-Y,0) {x=t;y=-W/2; }
border I1(t=-W/2,W/2) {x=S;y=t; }
border I2(t=S,S+Y) {x=t;y=W/2; }
border I3(t=W/2,-W/2) {x=S+Y;y=t; }
border I4(t=S+Y,S) {x=t;y=-W/2; }

// H1-H4 defines the injecting contact
//I1-I4 defines the collecting contact

mesh Th = buildmesh(G1(40)+G2(40)+G3(40)+G4(40)+H1(200)
             +H2(8)+H3(200)+H4(8)+I1(200)+I2(8)+I3(200)+I4(8));
fespace Vh(Th,P1);
 Vh p, w, pl, wl; // p the density of heavy holes, pl of light holes

//plot (Th, wait=true, value=true);

problem diffp(p,w,solver=LU) =
       int2d(Th)( dx(w)*dx(p) + dy(w)*dy(p))
       + int2d(Th)(w*p/L^2)
       - int2d(Th)(w*p0/L^2)
       + on(I1,p=p0) + on(I3,p=p0) + on(I4,p=p0) + on(I2,p=p0) 
       + on(H3,p=p1) + on(H1,p=p1) + on(H4,p=p1) + on(H2,p=p1);

problem diffpl(pl,wl,solver=LU) =
       int2d(Th)( dx(wl)*dx(pl) + dy(wl)*dy(pl))
       + int2d(Th)(wl*pl/Ll^2)
       - int2d(Th)(wl*pl0/Ll^2)
       + on(I1,pl=pl0) + on(I3,pl=pl0) + on(I4,pl=pl0) + on(I2,pl=pl0)
       + on(H3,pl=pl1) + on(H1,pl=pl1) + on(H4,pl=pl1) + on(H2,pl=pl1);

diffp;
diffpl;

//plot (Th, p, value=true, coef=1, wait=true);
//plot (Th, pl, value=true, coef=1, wait=true);
cout <<"holes  "<<p(0,0)<<"   "<<p(S,0)<< endl;

// calculate the hole current at the front of collecting contact
real Ih = int1d(Th,I1)( D*dx(p) );
real Ihl = int1d(Th,I1)( Dl*dx(pl) );
//real Ah = int1d(Th,I1)( 1.0 )*2.7e-4; // length of contact * thickness of InGaAs
cout <<" Heavy hole current, front edge   "<<-1.6e-19*Ih*2.7e-4<< endl;
cout <<" Light hole current, front edge   "<<-1.6e-19*Ihl*2.7e-4<< endl;

// calculate the y component of current at the front of collecting contact
// this should be essentially zero if the current is perpendicular to the edge
real Ihy = int1d(Th,I1)( D*dy(p) );
real Ihyl = int1d(Th,I1)( Dl*dy(pl) );
cout <<" y-comp of heavy hole current, front  "<<1.6e-19*Ihy*2.7e-4<< endl;
cout <<" y-comp of light hole current, front  "<<1.6e-19*Ihyl*2.7e-4<< endl;

// calculate the hole current at the top the collecting contact
real Ihty = int1d(Th,I2)( D*dy(p) );
real Ihtyl = int1d(Th,I2)( Dl*dy(pl) );
cout <<" y-comp heavy hole, top,    "<<1.6e-19*Ihty*2.7e-4<< endl;
cout <<" y-comp light hole, top     "<<1.6e-19*Ihtyl*2.7e-4<< endl;

// calculate the hole current at the collecting contact, backside
real Ih3 = int1d(Th,I3)( D*dx(p) );
real Ih3l = int1d(Th,I3)( Dl*dx(pl) );
//real Ah3 = int1d(Th,I3)( 1.0 )*2.7e-4; // length of contact * thickness of InGaAs
cout <<" heavy hole current, back    "<<1.6e-19*Ih3*2.7e-4<< endl;
cout <<" light hole current, back    "<<1.6e-19*Ih3l*2.7e-4<< endl;

// total current = sum of the four sides
cout <<"Total heavy hole current = "
      <<-1.6e-19*Ih*2.7e-4+2*1.6e-19*Ihty*2.7e-4+1.6e-19*Ih3*2.7e-4<< endl;
cout <<"Total light hole current = "
      <<-1.6e-19*Ihl*2.7e-4+2*1.6e-19*Ihtyl*2.7e-4+1.6e-19*Ih3l*2.7e-4<< endl;
cout <<"Total collected current  = "
      <<-1.6e-19*Ih*2.7e-4+2*1.6e-19*Ihty*2.7e-4+1.6e-19*Ih3*2.7e-4
      -1.6e-19*Ihl*2.7e-4+2*1.6e-19*Ihtyl*2.7e-4+1.6e-19*Ih3l*2.7e-4<< endl;

// x and y hole currents
Vh jpx = dx(p);
Vh jpy = dy(p);
//plot(p, [jpx,jpy], coef=8e-21, wait=1, value=true); 
//plot( jpx,  wait=1, value=true);

{
ofstream file("conc-two-hole-01.dat");
file << "# hole concentraton along x-axis, between contacts." << "\n";
file << "# x um     p_heavy    p_light" << "\n";
for (int i=0;i<=50;i++){
file << i*(S*1e4/50) << "     " << p(i*(S/50),0) <<"    "<<pl(i*(S/50),0)<< "\n";
    }
};

\end{lstlisting}
